\documentstyle [aps,preprint] {revtex}
\begin {document}
\draft {}
\title {Interference in the reaction $e^ + e^-\to\gamma\pi^ + \pi^-$ and
the  final state interaction.}

\author{N.N. Achasov
\thanks{ E-mail: achasov@math.nsc.ru} and  V.V. Gubin
\thanks{ E-mail: gubin@math.nsc.ru }}
\address{Laboratory of Theoretical Physics\\
S.L. Sobolev Institute for Mathematics\\
630090 Novosibisk 90,\  Russia}
\date{\today}
\maketitle
\begin{abstract}

We describe the interference between amplitudes 
$e^ + e^-\to\rho\to\gamma\pi^+\pi^-$ and 
$e^ + e^-\to\phi\to\gamma (f_0 + \sigma) \to\gamma\pi^ + \pi^-$, 
with regard to the phase of the  elastic $\pi\pi$ scattering background 
and mixing of the $f_0$ and $\sigma$ mesons. It is shown that the
Fermi-Watson theorem for the final state interaction  in the reaction
$e^ + e^-\to\rho\to\gamma\pi^ + \pi^-$ is not valid in the case of soft
photons, $\omega < 100$ MeV, in the $\phi$ meson region. The interference 
patterns in the spectrum of the photon energy differential cross-section 
and in the full cross-section as a function of the beam energy are obtained.

\end {abstract}


As is known, the problem of scalar $f_0$ and $a_0$ mesons is the central 
problem of the spectroscopy of light hadrons. Theoretical 
investigations have established that the study of the decays
$\phi\to\gamma f_0\to\gamma\pi\pi$ and $\phi\to\gamma a_0\to\gamma\pi\eta$ 
can shed  light on this problem 
\cite {{achasov-89},{achasov-95},{close},{molecule}}. In this paper 
we present results on
$\phi\rightarrow\gamma f_0\rightarrow\gamma\pi^ + \pi^-$ decay.
Experimentally, the radiative decays $\phi\rightarrow\gamma
f_0\rightarrow\gamma\pi^ + \pi^-$ are studied by observing the interference 
patterns at the $\phi$ meson peak in the reaction
 $e^ + e^-\to\gamma\pi^ + \pi^-$  which is currently studied in 
 the detector CMD-2 \cite {novo} at $e^ +e^-$-collaider VEPP-2M in
Novosibirsk and  will be  studied at the $\phi$-factory DA$\Phi$NE
when it comes into operation.

The analysis of the interference patterns in this reaction, 
by virtue of a large radiative background, is a rather complex problem 
which was actively considered in the current literature, 
see \cite {interference} and references therein.

It is necessary to note, that in paper \cite {interference} and in all
previous papers (see \cite {interference}) this problem was actually 
considered in the approximation of the single $f_0$ meson production,
 for example, as the $\rho$ meson. 
However, it is necessary to take into account 
that the $f_0$ meson is strongly coupled not only with the $\pi\pi$
and $K\bar K$ channels but also with other scalar resonances (like a 
$\sigma$ meson) and also with an elastic background.
 This fact, as we shall show below, 
deforms considerably the interference patterns both in the
spectrum of photons and in the full cross-section of the  reaction.

In view of this, we represent the analysis of interference patterns in the 
reaction $e^+e^-\to\gamma\pi^+\pi^-$ at the $\phi$ meson peak 
with regard to the phase of the elastic $\pi\pi$ scattering background 
 and  the $f_0$ and $\sigma$ meson mixing. As a reference point we consider 
experiment \cite {novo} and preliminary data obtained in this experiment.

Analyzing the phase relations between the amplitudes 
$\gamma\gamma^ * (s) \to\pi\pi$ and $\pi\pi\to\pi\pi$, we show that
 the Fermi-Watson theorem on the final state interaction in the 
 $e^ + e^-\to\gamma^ * (s)\to\rho\to\gamma\pi^ + \pi^-$ reaction 
is not valid, i.e. the phase of the amplitude 
$\gamma^ * (s) \to\gamma\pi\pi$ is not determined by the phase of $\pi\pi$
scattering amplitude. For soft  photons in the amplitude
 $\gamma^ * (s) \to\rho\to\gamma\pi\pi$ the Born term, i.e. 
the bremsstrahlung, is dominant, as the Low theorem requires, and  
the phase of amplitude in the elasticity region is determined by the
phase of the $\pi\pi$ scattering in theisovector vector channel 
( by the $\rho$ meson propagator ). At $4m_ {\pi} ^2/s, 
4m_ {\pi} ^2/m^2\ll1$, where $m$ is the invariant $\pi\pi$ mass, 
the amplitude is divided into two parts: i) the bremsstrahlung
 with the phase of the $\pi\pi$ scattering in isovector vector channel,
 which dominates in the soft photons region, $s\simeq m^2$, and ii) 
the amplitude of structural radiation,
 the phase of which in the elastic region
is determined by a sum of phase of the $\pi\pi$ scattering in the isovector
vector channel ($\delta^1_1 (E) $) and the phase of the $\pi\pi$ scattering in
isoscalar scalar (in our case) channel ($\delta^0_0 (m) $), the analog
of the Fermi-Watson theorem \cite{creutz}. This fact
leads to the shift of the interference pattern both in the full and in
the differential cross-section of the process.

In our case the classical Fermi-Watson theorem is faced with the problem of
soft photon radiation. The general  theoretical reasons for this case 
were adduced in \cite {creutz}, where there was noted,
that the Low theorem on a soft photon is not consistent with the analog 
of the Fermi-Watson theorem on the final state interaction. The way out of  
this situation the authors of Ref.\cite {creutz} see in separation of 
a photon energy spectrum into "soft" and "hard" parts  in each of
which the "proper" theorem holds. 
It is necessary to note that in the formal approach of \cite {creutz}
the separation of a photon spectrum into "soft" and "hard" parts looks 
 somewhat as a trick that reconciles these two theoretical conclusions.
In contrast \cite {creutz} we  consider a particular
representation of the amplitude of the reaction 
$e^ + e^-\to\gamma^ *\to\rho\to\gamma\pi^ + \pi^-$, see. Fig.1, 
and  demonstrate when the Fermi-Waston theorem holds, when the amplitude 
satisfies to the Low theorem and when it is broken up into two contributions: 
the bremsstrahlung and an amplitude satisfying to the analog of the 
Fermi-Watson theorem.

Let us consider the two-photon production of pions,
 $\gamma\gamma^ * (s) \to\pi\pi$,
below inelastic thresholds of the $\pi\pi$ scattering. In the case when
$s < 4m_ {\pi} ^2$ the pion interaction in the final state results in 
 a common phase of the  amplitude which equals the phase of the $\pi\pi$
 scattering.

Really, in the approximation, taking into account only two-pion
intermediate states, we assume that the amplitude of the $\pi\pi$
scattering ( see the last diagram in Fig.1(a) ) lies on the mass shell. Then
the full $s$-wave isoscalar amplitude of the process under interest is
determined by expression, see. Fig. 1
\begin {equation}
\label {treug} 
T=\frac {2} {3} \frac {e^2g_ {\rho\pi\pi}} {8\pi f_ {\rho}}
[\frac {m^2} {E^2-m^2}\frac {1-\rho_{\pi\pi} (m) ^2} {2\rho_{\pi\pi}(m)}
\lambda (m)+\frac{E^2}{m^2-E^2}+T_{\pi\pi}^{I=0}(m)g_{\pi\pi} (E,m)]
\frac {m_ {\rho} ^2} {D_ {\rho} (E^2)},
\end {equation}
where $m$ is invariant mass of $\pi\pi$ system, $\rho_{\pi\pi}(m) =\sqrt
{1-4m_{\pi}^2/m^2}$, $\lambda(m)=\ln[(1+\rho_{\pi\pi}(m))/(1-\rho_{\pi\pi}
(m))]$ and $1/D_{\rho}(E^2)$ is the $\rho$ meson propagator presented in
\cite{interference}. The function $g_{\pi\pi}(E,m)$ is determined 
by the triangular diagram \cite{achasov-89}.

When $E<2m_{\pi},\ m> 2m_ {\pi} $
\begin {eqnarray} 
&& g_ {\pi\pi}(E,m)=\frac{1}{\pi}\Biggl\{1+\frac{1-\rho_ {\pi\pi} ^2 (m)}
{\rho_ {\pi\pi} (E) ^2-\rho_ {\pi\pi}(m)^2}\Biggl
[\rho_ {\pi\pi} (m) (\lambda (m) -i\pi) + 2 | \rho_ {\pi\pi}(E)|\arctan(|
\rho_ {\pi\pi} (E) |) -\nonumber \\ 
&& -\pi |\rho_ {\pi\pi} (E) | -\frac {1}
{4} (1-\rho_ {\pi\pi} (E) ^2) \Biggl(\pi^2-2\pi\arctan (| \rho_ {\pi\pi} (E)
|) -\lambda (E) ^2 (\pi + i\lambda (m)) ^2\Biggr) \Biggr] \Biggr\},
\end {eqnarray}
when $E > 2m_ {\pi},\ m> 2m_ {\pi} $
\begin {eqnarray}
\label {funcgother} 
&& g_ {\pi\pi} (E, m) =\frac {1} {\pi}\Biggl\{1 + \frac
{1-\rho_{\pi\pi}^2(m)} {\rho_ {\pi\pi} (E) ^2-\rho_ {\pi\pi} (m) ^2} \Biggl
[\rho_ {\pi\pi}(m)(\lambda (m) -i\pi) - \nonumber \\ 
&& -\rho_{\pi\pi} (E)
(\lambda (E) -i\pi) -\frac {1} {4} (1-\rho_ {\pi\pi} (E) ^2) \Biggl ((\pi +
i\lambda (E)) ^2- (\pi + i\lambda (m)) ^2\Biggr) \Biggr] \Biggr\}. \end
{eqnarray}

We consider the parametrization of the $s$-wave amplitude of the $\pi\pi$
scattering below the inelastic thresholds in the following manner
 \begin {equation}
\label {eqpipi}
 T_ {\pi\pi} ^ {I=0} (m) =\frac {e^ {2i\delta^0_0 (m)} -1}
{2i\rho_ {\pi\pi} (m)}.
 \end {equation}
Substituting the amplitude (\ref {eqpipi}) in (\ref {treug})  one easily 
gets in the case $E < 2m_ {\pi} $ 
\begin {eqnarray}
&& T= \frac {2} {3}\frac {e^2g_ {\rho\pi\pi}} {8\pi f_ {\rho}} e^
 {i\delta^0_0 (m)} [(\frac {m^2}
{E^2-m^2} \frac {1-\rho_ {\pi\pi} (m) ^2} {2\rho_ {\pi\pi} (m)} \lambda (m)
-\frac {E^2} {E^2-m^2}) \cos (\delta^0_0 (m)) + \nonumber \\ 
&& + \sin(\delta^0_0 (m)) \frac {Re (g_ {\pi\pi} (E, m))}
 {\rho_ {\pi\pi} (m)}] \frac{m_ {\rho} ^2} {D_ {\rho} (E^2)}
\end {eqnarray}
and the Fermi-Watson theorem is valid, i.e. the phase of the 
$\gamma\gamma^ * (E) \to\pi\pi$ process is determined by the phase of the
$\pi\pi$ scattering. However, it is easy seen from (\ref {treug}) and
 (\ref {funcgother}) that the $\rho_ {\pi\pi} (E) $ and  $\lambda (E) $ 
terms in the imaginary part of the triangular diagram destroy 
the Fermi-Watson theorem in the case $E > 2m_ {\pi} $.

For soft photons ($E\simeq m$) the triangular diagram is proportional
to the factor ($E^2-m^2$), see (\ref {funcgother}), and in (\ref {treug})
the Born term, i.e. the bremsstrahlung, is dominant in the full agreement
with the Low theorem 
\footnote{ Actually, this conclusion does not depend on our 
assumption that the amplitude of the $\pi\pi$ scattering in the last diagram
in Fig.1(a) lies on a mass shell since this diagram vanishes in
the soft photon region due to the gauge invariance.}
and the amplitude phase is determinated by the phase of
the $\rho$ meson propagator.

In case when $E^2, m^2\gg4m_ {\pi} ^2$ the imaginary part of the
 triangular diagram 
\begin {equation} 
Im (g_ {\pi\pi} (E, m)) \simeq\frac {2m_ {\pi} ^2}
{m^2} (1 + \frac {m^2} {E^2-m^2} \ln\frac {m^2} {E^2})
\end {equation}
 is small since the imaginary parts from the  $E$ and $m$ channels 
 compensate each other. The amplitude of the process in this region  is 
broken up into two contributions: the bremsstrahlung which dominates in the
soft photon region and the amplitude satisfying the analog of the 
Fermi-Watson theorem \cite {creutz}:
\begin {equation}
\label {soft}
 T=\frac {2} {3} \frac {e^2g_ {\rho\pi\pi}} {8\pi f_ {\rho}}
[\frac {E^2} {m^2-E^2} + e^ {i\delta^0_0 (m)} \sin (\delta^0_0 (m)) Re (g_
{\pi\pi} (E, m))] \frac {m_ {\rho} ^2} {D_ {\rho} (E^2)}. 
\end {equation}

At $E > m> 2m_ {\pi} $ our amplitude describes the process 
$e^ + e^-\to\gamma^*\to \rho\to\gamma\pi^ + \pi^-$. In the soft photon 
region (energy of a photon $\omega < 100$ MeV) which we consider below, 
since $m_ {f_0} \simeq m_{\phi} $, the second term in (\ref {soft}) 
is negligible in comparison to the first one and it is negligible also
in comparison to the contribution of the $\phi$ meson, 
$e^+e^-\to\phi\to\gamma (f_0 + \sigma) \to \gamma\pi^ + \pi^-$, because of
the narrow width of the $\phi$ meson 
(smallness is proportional $\Gamma_ {\phi} m_ {\phi} /
 (m_ {\phi} ^2-m_ {\rho} ^2) \simeq1/100$).

The whole formalism for the description of the reaction under study $e^ +
e^-\to\phi\to\gamma f_0\to\gamma\pi^ + \pi^-$ was stated in \cite
{interference}, here we write out only the basic formulae modified by the
mixing of the $f_0$ and $\sigma$ mesons and by an elastic background of the
$\pi\pi$ scattering \cite {sigma}.

The amplitude of the $e^- (p_1) e^ + (p_2) \to\phi\to\gamma (f_0 + \sigma)
\to \gamma (q) \pi^ + (k_ +) \pi^- (k_-) $ reaction is written down
in the following way \cite{achasov-84,sigma}
\begin {equation}
M=e\bar u (p_1)\gamma^{\mu}u(p_2)\frac{em_{\phi}^2}{f_ {\phi}}\frac
{e^ {i\delta_B (m)}g(m^2)}{sD_ {\phi} (s)} (q^ {\mu} \frac {e (\gamma) p}{pq}
 -e (\gamma)^{\mu})\sum_{RR'}(g_{RK^+K^-}G^{-1}_{RR'}(m)
 g_{R'\pi^ + \pi^-})
 \label {amplituda}
 \end {equation}
where $s=E^2=p^2= (p_1 + p_2) ^2$, $t=m^2= (k_- + k_ +) ^2$, and the 
summation is over the two resonances $R=f_0,\sigma$.
 The definition of the function $g (m^2)$ and the $\phi$ meson propagator
$1/D_{\phi}(s)$ are given in \cite {sigma}.
 The data on the $\pi\pi$ scattering in the isoscalar  channel are 
parametrized as follows \cite {achasov-84,sigma}:
\begin {equation}
 T (\pi\pi\to\pi\pi) =\frac {\eta^0_0e^ {2i\delta^0_0 (m)} -1}
{2i\rho_ {\pi\pi} (m)}=\frac {e^ {2i\delta_B (m)} -1} {2i\rho_ {\pi\pi} (m)}
+ e^ {2i\delta_B (m)} T^ {res} _ {\pi\pi} (m),
\end {equation}
where $\delta^0_0 (m) =\delta_B (m) + \delta_ {res} (m) $ and 
\begin{equation}
\label {amplitudapipi}
 T^{res}_{\pi\pi} (m) =\sum_ {RR'} \frac {g_
{R\pi\pi} g_ {R'\pi\pi}} {16\pi} G^ {-1} _ {RR'} (m).
\end {equation}
The phase of the elastic background $\delta_B (m) $ is taken in the form
$\delta_B=\theta\rho_ {\pi\pi} (m) $, where $\theta\simeq60^ {\circ}$.
The matrix of the inverse propagator $G_ {RR'} (m)$ is presented 
in \cite {achasov-84,sigma}. The amplitude of the $\pi\pi\to K\bar K$ process
is 
\begin {equation}
\label {amplppkk}
T_ {K\bar K} =e^{i\delta_B (m)}\sum_{RR'}\frac{g_{R\pi\pi}g_{R'K\bar K}}
{16\pi}G^{-1}_{RR'}(m).
\end {equation}

Thus, for the differential cross-section of the signal we obtain the
following  expression
\begin {equation}
\label {signal}
 \frac {d\sigma_ {\phi}} {d\omega} =\frac {\alpha^2\omega} {8\pi
s^2} \left (\frac {m_ {\phi} ^2} {f_ {\phi}} \right) ^2\frac {| g (m^2) | ^2}
{| D_ {\phi} (s) | ^2} \sqrt {1-\frac {4m_ {\pi} ^2} {m^2}} (a + \frac {a^3}
{3}) b \left | \sum_ {RR'} (g_ {RK^ + K^-} G^ {-1} _ {RR'} (m) g_ {R'\pi^ +
\pi^-}) \right | ^2 H_ {rad} (s,\omega_ {min}),
\end {equation}
where $\omega= |\vec q|$ is energy of the photon, $a$ is the cut on
 $\theta_{\gamma} $, the angle between the photon momentum and
the electron beam in the center-of-mass  frame of the reaction,
$-a\leq\cos\theta_ {\gamma} \leq a$ and $b$ is the cut  on
$\theta_ {\pi\gamma}$, the angle between the photon and pion momenta 
in the dipion rest frame, $-b\leq\cos\theta_{\pi\gamma}\leq b$.

The function $H_{rad}(s,\omega_ {min}) $ takes into account the radiative
corrections,  the contribution of which reduces the cross-section 
by 20\%,  see \cite {{interference},{sigma}}.

As was shown in the previous papers, see \cite {interference}, the main
background to process under study has come from  the  initial electron
radiation  and the radiation from the final pions. The background from 
non-resonant  invariant mass of the $\pi^+\pi^-$  system processes was 
estimated in \cite {interference,sigma} and in our region of the spectrum,
 $20 < \omega < 100$ MeV, was found to be negligible.

The differential cross-sections for the background processes connected 
with the final state radiation and the  initial one have the following 
expressions
\begin {eqnarray}
\label {fonrho}
&&\frac{d\sigma_f} {d\omega} =2\sigma_0 (s) \frac {1} {\sqrt
{s}} F (x, a, b) |1-\frac {3\Gamma (\phi\to e^ + e^-) \sqrt {s}} {\alpha D_
{\phi} (s)} |^2 H_ {rad} (s,\omega_ {min}) \\ \nonumber 
&&\frac{d\sigma_i}{d\omega}=2\sigma_0 (m^2) \frac {1} {\sqrt {s}}H(x, a, b)
|1-\frac{3\Gamma(\phi\to e^+e^-) m} {\alpha D_ {\phi} (m^2)}|^2
 H_ {rad} (m,\omega_{min}),
\end {eqnarray}
where $x=2\omega/\sqrt {s}$. The functions $F(x,a,b)$, $H(x,a,b)$ and the 
cross-section $\sigma_0(s)$ of the $e^ + e^-\to\pi^ + \pi^-$ process
are presented in \cite{interference}.

The interference of the amplitude (\ref {amplituda}) with the amplitude of
the final pions radiation is 
\begin {eqnarray} 
&&\frac{d\sigma_ {int}}{d\omega}=\sqrt{\frac{3}{2}}\frac{\alpha^3}{s\sqrt{s}}
\left(\frac{g_{\rho\pi\pi}}{f_ {\rho} f_ {\phi}}\right) Re\Biggl[\frac 
{m_ {\phi} ^2m_{\rho} ^2g (m^2) e^{i\delta_B (m)}}{\sqrt{4\pi\alpha} 
D_ {\phi} D^ * _{\rho}} (1-\frac {3\Gamma (\phi\to e^ + e^-) \sqrt {s}}
 {\alpha D^ * _ {\phi}(s)}) \times \\ \nonumber 
&& \times\left(\sum_ {RR'} g_ {RK^ + K^-} G^ {-1} _{RR'}g_ {R'\pi^ + \pi^-}
 \right) \Biggr] \Biggl\{f (x) + \frac {\xi} {2}\ln\frac {1-f (x)}
 {1 + f (x)} \Biggr\} (a + \frac {a^3} {3}) H_ {rad}(s,\omega_ {min}),
\end {eqnarray}
where $f(x)=b\sqrt {1-\frac {\xi} {1-x}} $ and $\xi=4m_ {\pi} ^2/s$.

The total differential cross-section is
$d\sigma/d\omega=d\sigma_ {\phi}/d\omega
+d\sigma_i/d\omega + d\sigma_f/d\omega\pm d\sigma_ {int} /d\omega$.

The fitting of the $\pi\pi$ scattering data shows that a number of parameters 
describe well the  $\pi\pi$ data in the region of interest $0.7<m<1.8$ GeV,
 see \cite {sigma}.

By way of an illustration we present the interference patterns for 
two sets of parameters, see. Fig. 2, for the narrow  
and relatively wide $f_0$ resonance:

i) $m_ {f_0} =980$ MeV, $m_ {\sigma} =1.47$, $R=8$, $g^2_ {f_0K^ + K^-}
/4\pi=2.25\ GeV^2$, $g_{\sigma\pi\pi} ^2/4\pi=1.76\ GeV^2$, 
$g_ {\sigma K\bar K} =0$, $\theta=60^ {\circ} $, 
the subtraction constant for non-diagonal elements of the matrix of the
inverse propagator $C_ {f_0\sigma} =-0.31$ GeV \cite{sigma},
 the effective width of the  $f_0$ meson $\Gamma_ {eff} =0.025$ GeV. 
The branching ratios
of the decay for this set of parameters are:
$BR (\phi\to\gamma f_0\to\gamma\pi\pi,\omega<250\ MeV)=3.36\cdot10^{-4}$,
$BR (\phi\to\gamma (f_0+\sigma)\to\gamma\pi\pi, \omega<250\ MeV)=
2.74\cdot10^ {-4} $ and
$BR (\phi\to\gamma f_0\to\gamma\pi\pi,\omega < 100\ MeV) = 0.79\cdot10^ {-4}$,
$BR (\phi\to\gamma (f_0 + \sigma) \to\gamma\pi\pi,\omega < 100\ MeV) =
1.0\cdot10^ {-4} $.

ii) $m_ {f_0} =985$ MeV, $m_ {\sigma} =1.38$, $R=2$, $g^2_ {f_0K^ + K^-}
/4\pi=1.47\ GeV^2$, $g_ {\sigma\pi\pi} ^2/4\pi=1.76\ GeV^2$,
 $g_ {\sigma K\bar K} =0$, $\theta=45^ {\circ} $, 
the subtraction constant for non-diagonal elements of the matrix of the 
inverse propagator $C_ {f_0\sigma} =-0.31$ GeV \cite{sigma},
 the effective width
of the  $f_0$ meson $\Gamma_ {eff} =0.085$ GeV. The branching ratios
of the decay for this set of parameters are:
$BR (\phi\to\gamma f_0\to\gamma\pi\pi, \omega < 250\ MeV)=2.32\cdot10^{-4}$,
$BR (\phi\to\gamma (f_0 + \sigma) \to\gamma\pi\pi, \omega < 250\ MeV) =
1.67\cdot10^ {-4} $ and
$BR (\phi\to\gamma f_0\to\gamma\pi\pi,\omega < 100\ MeV)=0.68\cdot10^ {-4}$,
$BR (\phi\to\gamma (f_0 + \sigma) \to\gamma\pi\pi,\omega < 100\ MeV) =
0.59\cdot10^ {-4} $.

The interference patterns in the full cross-section of the reaction $e^ +
e^-\to\gamma\pi^ + \pi^-$, $\sigma_ {\phi} \pm\sigma_ {int} + \sigma_f +
\sigma_i$, by energy of beams at the $\phi$ meson region for both
sets of parameters are shown in Fig.2(a).  Being guided by \cite {novo},
we have chosen the angular cuts  $a=0,66$ and $b=0.955$ which suppress
the contribution of the initial state radiation by a factor of nine.
 However, despite the strong suppression, the initial state radiation
remains dominant and is  about $\frac{2}{3}$ of the total background. 
The photon energy is in the interval $20 < \omega < 100$ MeV.

The interference pattern in the photon  spectrum  $d\sigma_ {\phi} /d\omega
\pm\sigma_ {int} /d\omega$ at the $\phi$ meson peak is shown in Fig.2(b).

This  analysis of the interference patterns in reaction $e^ +
e^-\to\gamma\pi^ + \pi^-$ shows that the study of this reaction is 
an interesting and rather complex problem. In this contribution we have shown
that the analysis of the interference patterns in the
$e^ + e^-\to\gamma\pi^ + \pi^-$ reaction needs to be performed along with 
the analysis of reactions $\pi\pi\to\pi\pi$ and $\pi\pi\to K\bar K$.
The experimental study of the $e^ + e^-\to\gamma\pi^ + \pi^-$ 
reaction can  obviously enrich our knowledge about nontrivial
manifestations  of the strong interactions and the features 
of the $f_0$ meson.

We thank  J.A. Thompson  for help with English.

This work was supported by INTAS, grant INTAS-94-3986.

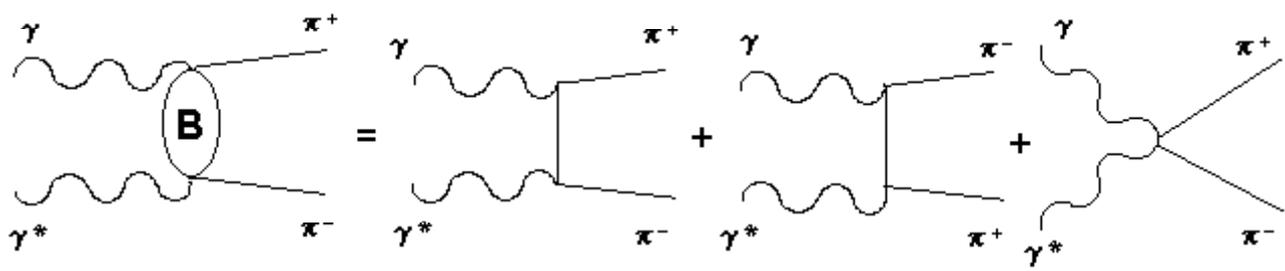
\begin {figure}
\caption {The diagrams of the model.}
 \end {figure}

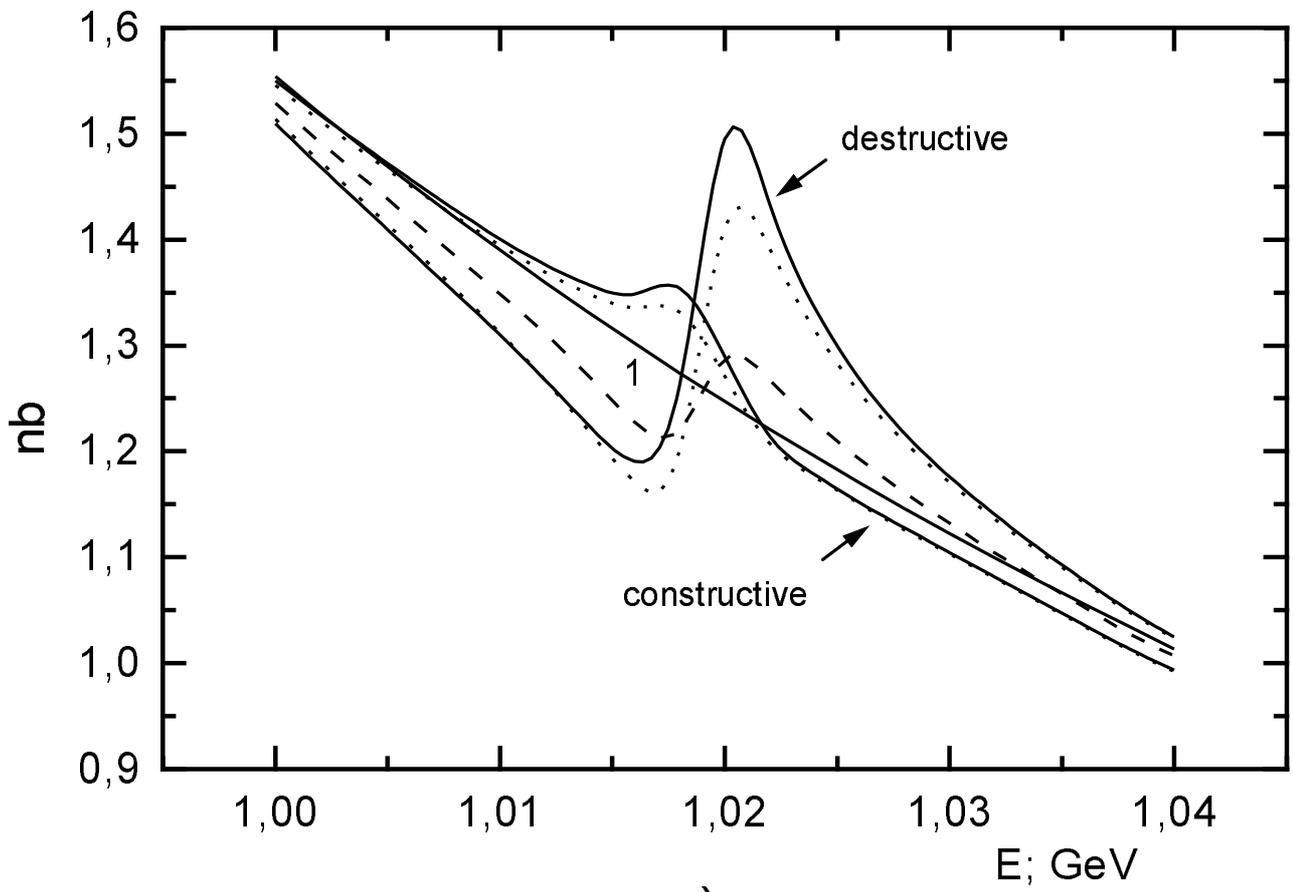
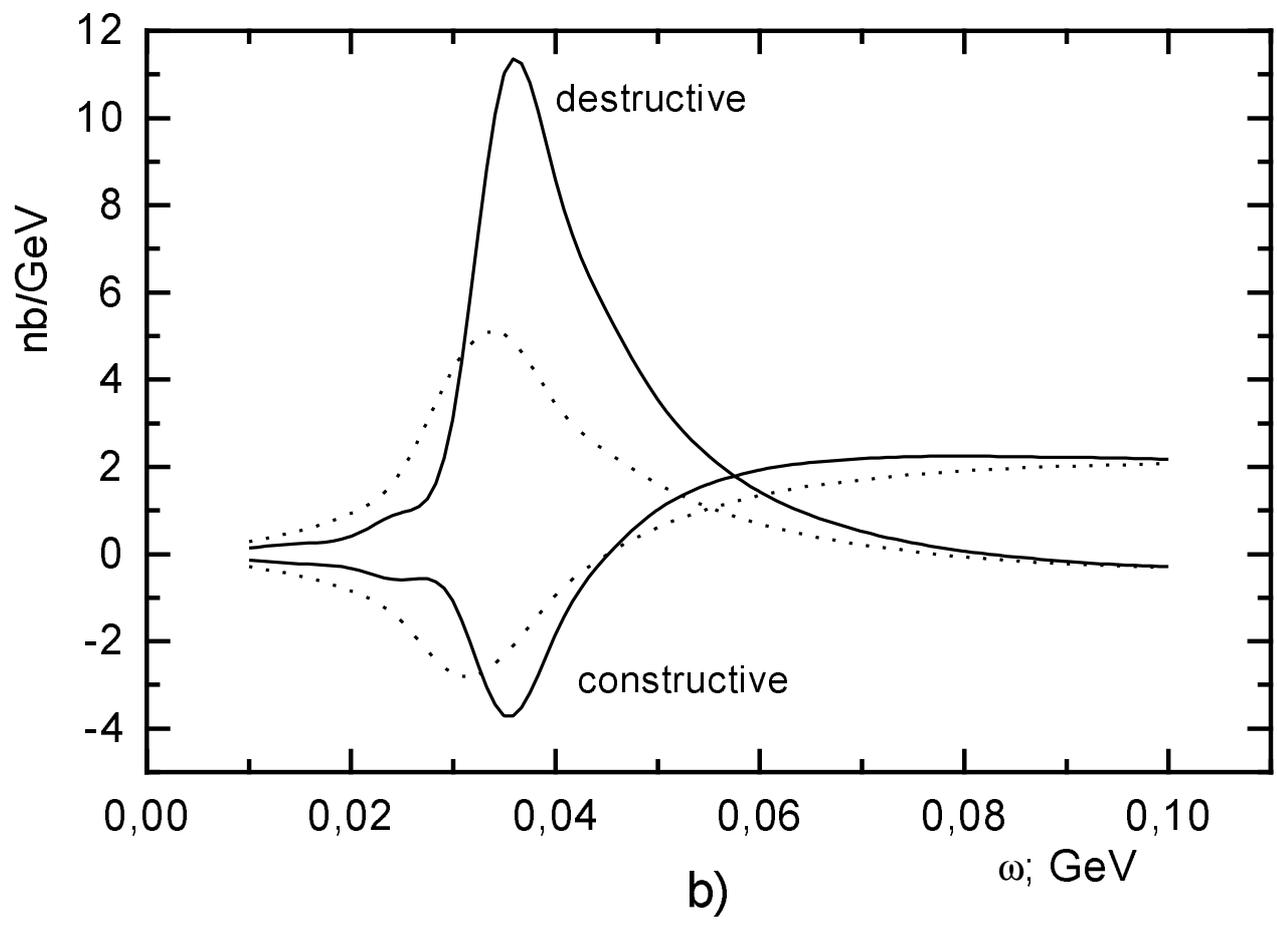
\begin {figure}
\caption {(a) The interference pattern in the total cross-section.
 The solid line 1 corresponds to the pure background and the dashed line 
is the background with the $\phi-\rho$ transition, see [6].
The solid lines for
destructive and constructive interferences correspond to 
$\Gamma_ {eff} =0.025$ GeV. The dotted lines for destructive and constructive
interferences correspond to $\Gamma_ {eff} =0.085$ GeV, see the text.
(b) The interference pattern in the spectrum of photons. The solid lines
 correspond to the narrow $f_0$ resonance, $\Gamma_ {eff} =0.025$ GeV and
the  dotted lines correspond to the relatively wide resonance, 
$\Gamma_ {eff} =0.085$ GeV.}
\end {figure}


\begin {references}
\bibitem {achasov-89}
N.N. Achasov, V.N. Ivanchenko, Nucl. Phys. {\bf B315}, 465 (1989);
Report No. INP 87-129, 1987 (unpublished).
\bibitem {achasov-95}
N.N. Achasov in {\it The second  DA$\Phi$NE Phsics Handbook},
 Eds L. Maiani, G. Pancheri, N. Paver, dei Laboratory Nazionali di Frascati,
 Frascati, Italy, May 1995, Vol.II, p.671.
\bibitem {close}
F.E. Close, N. Isgur and S. Kumano, Nucl. Phys. {\bf B389} 513 (1993).\\
N. Brown and F. Close in {\it The second  DA$\Phi$NE Phsics Handbook},
 Eds L. Maiani, G. Pancheri, N. Paver, dei Laboratory Nazionali di Frascati,
 Frascati, Italy, May 1995, Vol.II, p.649.
\bibitem {molecule}
N.N. Achasov, V.V. Gubin and  V.I. Shevchenko, Phys. Rev. D, to be published, 
(hep-ph/9605245).
\bibitem {novo}
R.R.  Akhmetshin et al, Budker report No. INP 95-62, 1995. 
\bibitem{interference}
N.N. Achasov, V.V. Gubin and E.P. Solodov, Phys. Rev.{\bf D55} 2672 (1997)
\bibitem {sigma}
N.N. Achasov and  V.V. Gubin, ( hep-ph/9703367 ).
 \bibitem {creutz}
M.J. Creutz and M.B. Einhorn, Phys. Rev. {\bf D1} 2537 (1970).
 \bibitem{achasov-84}
N.N. Achasov, S.A. Devyanin and G.N. Shestakov, Usp. Fiz. Nauk {\bf 142}
 361 (1984)[Sov. Phys. Usp.{\bf 27}, 161 (1984)]
\bibitem {bukin}
A.D. Bukin et al., Yad.Fiz. {\bf 27} 985 (1978) [Sov. J. Nucl. Phys.
 {\bf 27}, 521 (1978)]
 \end {references}
\end {document}
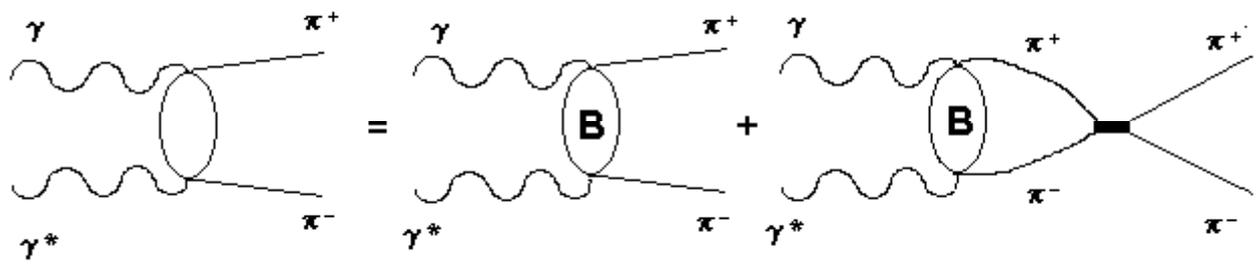
